\documentclass[12pt,drftcls,onecolumn]{IEEEtran}

\usepackage{cite}
\usepackage{graphicx}
\usepackage{setspace}
\usepackage{amsfonts}
\usepackage{amsmath}
\usepackage{textcomp}
\usepackage{algorithm}
\usepackage{algorithmic}
\usepackage{amsmath}

\usepackage{amsmath}
\usepackage{graphicx}
\include{psfig}
\usepackage{epsfig}
\usepackage{array}
\usepackage{psfrag}
\usepackage{amssymb}
\usepackage{color}
\usepackage{pstricks,pst-node,pst-text,pst-3d,pst-plot}

\usepackage{cite}

\def\nn{{\nonumber}}

\def\bx{{\bf x}}
\def\bnu{{\boldsymbol \nu}}

\def\bm{{\bf m}}

\def\bz{{\bf z}}

\def\by{{\bf y}}

\def\bI{{\bf I}}

\def\by{{\bf y}}

\def\bI{{\bf I}}

\def\bx{{\bf x}}
\def\by{{\bf y}}

\def\bN{{\bf N}}

\def\pr{{\rm Pr}}
\def\EN {\mathbb{E}\{\bN\}}
\def\ENd {\mathbb{E}\{\bN_d\}}
\def \ms {\bm_{\rm s}}

\def \jdh {J_D^{-h}}

\def\beta{{\boldsymbol \eta}}

\def\bnu{{\bf v}}

\title{Mobility Diversity in Mobile Wireless Networks}
\author{Veria Havary-Nassab$^*$, Shahram Shahbazpanahi$^*$\thanks{S.~Shahbazpanahi is the corresponding author. V.~Havary-Nassab and S.~Valaee are with the Department of Electrical and Computer Engineering, University of Toronto, 10 King's College Road, Toronto, ON, Canada, M5S 3G4, emails:
\texttt{veria.havarynassab@mail.utoronto.ca, valaee@comm .utoronto.ca}
phone:  +1  (416) 946-8032, fax: +1 (416) 978-4425. S.~Shahbazpanahi is with the Department of Electrical, Computer, and Software Engineering, University of Ontario Institute of Technology, 2000 Simcoe Street North, Oshawa, ON, Canada, L1H 7K4, e-mail:
\texttt{shahram.shahbazpanahi@uoit.ca},
phone: +1 (905) 721-8668, fax: +1 (905) 721-3370.}, Shahrokh Valaee$^*$}
\begin{document}

\baselineskip 9mm
\maketitle

\begin{abstract}

We introduce the novel concept of \emph{mobility diversity} for mobile sensor or
communication networks as the diversity introduced by transmitting data over different topologies
of the network. We show how node mobility can provide diversity by changing the topology of the
network. More specifically, we
consider a mobile network of a sensor node and a number of sink nodes which are all moving randomly according to different Wiener process mobility models. Assuming that the network topology evolves with time and assuming that the connectivity of the sensor node to at least one sink node is needed for successful communication, we calculate three performance measures for this network, i) the expected number of time instants, where the sensor node is connected to at least one sink node, ii) the probability of outage, being the probability that no sink node is in the vicinity of the sensor node during the observation interval, and finally, iii) the maximum number of consequent failures in the communication. Our theoretical and numerical analysis show that increasing the mobility parameter
of the sensor node increases the average number of successful transmissions, decreases the probability of outage, and reduces the maximum delay in the senor-sink communication.

\end{abstract}

\newpage

\section{Introduction}

Diversity schemes benefit from the fact that transmitting  a signal over multiple uncorrelated channels will enhance communication reliability compared to the traditional schemes, where one channel is used for signal transmission. The term \emph{diversity gain} is therefore adopted for such schemes to emphasize that the improvement achieved in communication reliabilityand/or the enhancement in the data rate comes from utilizing \emph{different} (statistically independent) possible communication channels. Therefore, compared to the case of communicating through a single channel, sending multiple copies of the information over multiple channels increases the probability of successful communication.

Time, frequency,
spatial, user cooperation, and multi-user
diversity techniques are among several diversity schemes introduced and studied in the literature. In frequency (time) diversity schemes
\cite{timediversity1961,timediversity1965,timediversity2009,OFDM1,OFDM2},
multiple copies of the same data are transmitted over different
frequency bands (in different time intervals). The distances between these frequency bands (or  those between transmission time intervals) are chosen large enough such that the fading
characteristics  of individual channels are statistically
independent.

When data
is sent over different propagation paths, spatial diversity is exploited \cite{spatial2001,spatial2004,spatial2004-2,spatial2007}. This can be implemented by using transmit or receive antenna diversity \cite{Antenna}. Several copies of the signal are
sent over spatially  distanced antennas so that each signal travels a different path. Multiple input multiple output (MIMO) communication systems are very well studied schemes that benefit from the spatial diversity achieved by equipping the receiver and the transmitter with multiple antennas \cite{tarokh1,tarokh2}.

A user-cooperation diversity scheme is defined in the context of cooperative communication systems, where
different nodes share their resources to establish a
distributed MIMO system. Indeed in such systems, each node acts as one antennas for  another node, thereby helping each other in
transmitting and receiving data
\cite{Laneman2003,Laneman2004,Larsson2005}. A node which needs to transmit a message to its corresponding destination, first shares its
information with its neighboring node(s). After this local message exchange, all the nodes which have received the message,
collectively transmit the signal to the intended destination, thereby establishing a distributed MIMO, and thus achieving user-cooperation diversity.

As discussed in the above diversity schemes, diversity gain is achieved when information is transmitted over statistically independent communication channels. In this paper, we introduce a novel
diversity scheme, namely \emph{mobility diversity}. We define mobility diversity in the
context of mobile networks but it can be readily extended to any network with evolving topology. A data exchange method is said to
be using mobility diversity if the information intended to be exchanged between the network nodes, is
transmitted over different network topologies. The evolution of the network topology is a
result of the mobility of the network nodes. This evolving network topology justifies the terminology mobility
diversity.

The following analogy can be established with the current diversity
schemes: the different communication channels available
for information exchange in traditional diversity schemes are
analogous to different topologies of a mobile network. In each
topology, due to communication range limitations and possible obstacles, a certain node may communicate with its corresponding destination node only through
 several hops. In this case,  the data transmitted by such nodes will either be lost or arrive at the destination with long delays. However, as the nodes are mobile, the network topology is constantly changing. Hence, in subsequent transmissions, the network will have a different topology and those ``unlucky" nodes could
be in a direct contact with, or a few hops away from their corresponding destinations. This means that in  upcoming topologies, such nodes might be able to communicate their messages easily
with acceptable delay or with tolerably low error rates. The receiver at the destination node, will then combine the received signals over different topologies to decode the message with a high reliability.

Recently, mobility in wireless sensor networks has been
addressed in the literature. Different network mobility
models have been studied and the effects of the mobility on the performance measures of the network, such as the capacity and delay, have been
analyzed \cite{Mobility1,Mobility2,Mobility4}. It
has been proven that, for a network with growing number of nodes in a limited area and in the presence of interference, mobility can enhance the performance of a mobile ad-hoc network.

In \cite{Mobility1}, the nodes of a mobile network are assumed to be located randomly on a sphere of unit radius. Each node intends to communicate to another randomly chosen node on the sphere. Nodes can move randomly along their one dimensional great circle on the unit sphere. The performance of the network is measured in terms of throughput per source-destination pair. It is shown in \cite{Mobility1} that such a restricted mobility model enhances the performance of the network. More specifically, in this scenario, the throughput per source-destination pair can be kept constant when the size of the network grows; while in a stationary network, the throughput is shown to be decreasing by increasing the number of nodes.

The delay associated with the increase of the network size has been studied in \cite{Mobility2}. It has been shown that in both random walk and restricted mobility models, the delay increases with the increase of the network size. This delay factor is higher than that of a fixed (stationary) network.

The authors of  \cite{Mobility4} have studied the throughput-delay tradeoff in a multicast scenario, where each node intends to broadcast its data to several nodes. The throughput-delay tradeoff has been derived for one- and two-hop communication scenarios. It has also been shown that if the number of receiving nodes grows linearly with the number of transmitting nodes, mobility does not enhance the throughput, compared to the stationary network scenario.

\textcolor[rgb]{0.00,0.00,1.00}{Such studies indicate that mobility can be exploited in favour of the performance measures of a mobile communication network. The approaches discussed above, illustrate that mobility can compensate the degradation encountered due to the growth of the node population in return for a larger delay. However, to the best of our knowledge, the effects of mobility on a network with a fixed (and probably small) number of nodes, have not been studied.}

{In this paper, we study the problem of the effect of mobility on the coverage status of a sensor node in a short period of time between its sensing instants (transient coverage status) when it is moving in a mobile network of sink nodes. The sensor senses the phenomenon of interest periodically and tries to exchange its data with a close-by sink node right after the sensing. If due to the evolving topology of the network, the sensor node is not able to communicate with a sink node at the sensing moment, it keeps repeating its attempts in the time slot between the two sensing instants hoping that mobility will make communication (or coverage) possible in one (or some) of the attempts. We study the contribution of mobility on the ability of the sensor node to communicate its information within the limited time period till the next sensing in a probabilistic framework.}

 {The main differences between the results we present in this paper and the studies reviewed above can be summarized as follows:}
\begin{itemize}
\item {The scenarios studied in this paper and the mobility model adopted are quite different (and much more practical) compared to that of \cite{Mobility1,Mobility2,Mobility3}. The Wiener mobility model adopted in this study can be tailored to many real world applications such as vehicular communication or crowd motion (in crowd sensing). The model used in the above mentioned studies, either the one dimensional motion on the great circle of a sphere or random motion inside a unit circle have limited applications and are more suitable for the abstract study of the effect of mobility.}
\item {The focus of the studies presented in \cite{Mobility1,Mobility2,Mobility4} is on the effect of mobility in compensating the degradation in the network capacity due to the growth of the population of the nodes in a restricted area. The asymptotic behaviour of the network capacity has been compared in the stationary and mobile scenarios. In our study, we consider the contribution of mobility on individual nodes and in a configuration with a finite and even small number of communicating nodes. We are looking at the transient effect of mobility on the coverage status of a sensor node in a mobile sensor network within a limited time period.}
    \item {While the problem at hand in \cite{Mobility2} for instance, is communicating a number of packets among a network through a two-hop scenario and the contribution of mobility examined in an aggregate fashion, in our studies, the focus is on the topology and coverage problem and the benefit that each single node can experience if it addresses mobility properly.}
        \item { In the following paper, the effect of mobility on the transient coverage status of a mobile node in a mobile network is studied. In other words, the problem looks at a situation in which a sensor node has a limited amount of time to convey its sensed data to the server through the sink nodes. The contribution of the mobility to the coverage status of the sensor node within that limited time is considered. This problem has many applications and has not been addressed by the results of the previous studies.}
\item {In \cite{Mobility2} and \cite{Mobility3}, the network capacity considered is an aggregate criteria for the total performance of the network. This is useful in an ad-hoc network where all the nodes have packets to be sent to some (randomly chosen) other nodes. The effect of mobility on the individual experience of each node is not therefore studied. This is specially important in sensor networks compared to ad-hoc networks where the information of each sensor is critical to the proper performance of the system. }
\end{itemize}

In this paper, we investigate the effect of mobility on the communication performance of a sensor network with \emph{random and finite} number of nodes. Adopting a Wiener process to model the node mobility, we show that in a network with high node mobility, any sensor node has more opportunity, compared to a node in a network with stationary nodes, to communicate its data to sink nodes (which are in charge of collecting data from sensor nodes). Hence, higher node mobility results in more reliability and/or leads to higher capacity in the link between a sensor and sink nodes.

In \cite{havaryPIMRC}, we studied the effect of mobility on the average number of opportunities, a mobile sensor node will have, to communicate with one of several sink nodes which are also mobile.
This paper expands the primary results published in \cite{havaryPIMRC} by studying the effects of mobility on two other network performance parameters, namely the probability of outage and the longest transmission delay.
 In the next section, considering a one-dimensional mobility model, we discuss the underlying scenario, present the corresponding data model, and introduce  the performance parameters to be investigated in the paper. In Section~III, the first parameter, namely the average number of time instants, where a sensor node is in the communication range of a sink node, will be introduced and it is derived based on our data model. In Section~IV, we mathematically prove that this average number of time instants is an increasing function of the node mobility parameter of the sensor node, meaning that the node mobility will enhance the performance of the network, measured as the average number of communication opportunities for the sensor node. In Section~V, we introduce the second performance parameter as the outage probability and use our data model to derive it in terms of the node mobility parameter. We  also prove in this section that the outage probability is an increasing function of the mobility parameter of the sensor node. Section~V also discusses the third performance parameter, the longest number of consecutive time instances, where the source is not in the communication range of any sink node. This parameter is related to the maximum delay that can occur in communication from a sensor node to the sink nodes. We propose an approach to derive this parameter based on our data model. Section~VI proposes a method to extend the  studied one-dimensional scenario to a two-dimensional configuration and shows that the mobility has similar effects on the aforementioned performance metrics in a two-dimensional mobility model. Section~VII presents our numerical evaluations and Section~8 concludes the paper.

\section{Problem Setup and Data Model}

Consider a wireless sensor network with a number of sensor nodes and a few sink nodes. The sensor nodes sense their environment and transmit their measured data to the sink nodes. The sink nodes  are responsible to send the data to a fusion center for further processing. Communication between the sink nodes and the sensor nodes happens in a single hop. Each sensor node is able to transmit its data to a sink node (and ultimately to a fusion center), only if it is within the distance $d$ (called communication range) of that sink node. This model is similar to the model used in \cite{Mobility1} with some subtle differences. In this paper we assume that if a sensor node is in the communication range of a sink node, they can communicate regardless of the possible interference in the channel. This model allows us to study the effects of the node mobility  on different network performance metrics, without being concerned about the interference. {Interference can be mitigated by assigning a frequency band to the sensor network for communication and program the nodes to utilize the available channels in an ad-hoc manner as explained in \cite{havaryicassp} for instance.}

In case of a stationary network, the sensor  and the sink nodes are deployed properly so that each sensor node can connect to at least one sink node. However, in a mobile wireless sensor network, mobility will change the node-to-node distances over time, thereby resulting in a time-varying network topology.

Let us define the network connectivity graph based on the distances of the network nodes. Each sensor or sink node is represented by a vertex in the graph. Two vertices are connected with an edge if their distance is less than the communication range $d$. In such a graph, a sensor node is able to successfully communicate its sensed data if it is connected to at least one sink node. Therefore, the topology of the network determines the overall performance of the communication scheme. This implies that the node mobility affects the communication performance as it continuously changes the connectivity graph of the network. A sensor node that is not connected to any sink node in the network at a particular time instance, has a chance to be in the proximity of a sink node in the next time instance due to its own mobility and/or due to the mobility of the sink nodes.

Depending on the scenario, a sensor node can exploit the mobility and the evolving topology, in one of the two following ways. If the sensor transmits the same data over different time instances, it can exploit the transmission diversity offered by statistically independent channels. In such a case, the fusion center can combine the information received over different topologies to extract the transmitted data with a higher reliability. In this scenario, having more time instants in which the sensor node has a connection to a sink node, results in a higher communication reliability. The gain achieved in such a scheme is commonly called ``diversity gain''. Alternatively, the sensor can use each time instant, in which the sensor node has a connection to a sink node, to transmit new data. In such a scenario, a larger number of connected time instants results in higher data rates in the network. The gain achieved here is called ``multiplexing gain''.

The sensor may adopt one of the two scenarios or a mixture of both to exploit the communication opportunities that occur due to the mobility of the nodes. However, it has been shown that there is a fundamental trade-off between the diversity gain and the multiplexing gain \cite{TSEDavid}. The sensor can choose how much of each gain it needs based on this trade-off  and may adopt a scheme that achieves certain diversity and multiplexing gains. In a network with nodes moving randomly, the number of communication opportunities that a sensor node can benefit from during a given time interval, will be a random variable. Therefore, a higher expected value of the number of such communication opportunities translates into having a more reliable communication and/or a higher throughput.

{To study the concept of mobility diversity, consider the case where a single sensor node periodically senses the environment every $T$ seconds and is required to send its data to a fusion center through one of the many sink nodes available in the environment. Each sink node collects the information from the sensor node only if the sensor node lies within a distance $d$ of that sink node, thereby allowing a successful link to be established between them.}

{The sensor node attempts to transmit its data right after each sensing. If the topology of the network at that time instant allows the connection between the sensor node and any of the available sink nodes, the communication is successful and the sensor can wait for the next sensing. However, if no sink node is in the vicinity of the sensor node at the sensing time, the sensor adopts the following scheme to enhance its communication reliability.}

{The sensor divides the $T$ seconds till the next sensing into $n$ time instants $\{t_k\}_{k=1}^{n}$ assuming that $t=0$ is the current sensing time at which the sensor was not able to communicate to any sink node. The communication will be therefore attempted $n$ times on these $n$ time instances to achieve a successful communication and to increase the reliability of the link. The time instants are chosen based on the average mobility of the network so that a topology change is probable from $t_i$ to $t_{i+1}$.
}

We further assume that the sensor node and the sink nodes are mobile along the $x$ axis and that different nodes move independently on this axis. Although the one-dimensional mobility assumption may not always be realistic, this assumption is commonly used in assessing the effects of mobility on the performance of wireless networks (see \cite{Mobility1}, for instance). This is because the analysis based on one-dimensional mobility can be readily extended to 2 or 3 dimensions. Furthermore, our analysis can be extended to the case of multiple sensors if the movements of sensor nodes are independent. Therefore, the one-dimensional single sensor case can be used as a basis to study more general mobility scenarios.


To model the mobility of the network nodes, we describe the node locations as Wiener random processes. Using Wiener process for modeling the time-varying node locations is a common practice in applied mathematics and engineering. {Wiener process is a good model for random motion of a physical object whose displacement between the times $t_1$ and $t_2$ goes to zero when $t_2 - t_1$ goes to zero. In other words, objects modeled by Wiener motion, can not disappear at a point and reappear in a distant point in no time. A drifted Wiener process, discussed in Section~\ref{app} can be adopted to model the motion of vehicles in a single lane of a highway. It is a suitable process to model the motion of nodes whose mobilities are random and independent of each other either in its continuous form or in modeling the mobility within a grid.}

In a sensor network, a Wiener model represents the  $x$ coordinate of a node as a Gaussian random variable with a certain mean and a variance which increases with time.

Assume that the time-varying $x$ coordinate of the sensor node, and that of the $j$th sink node are denoted, respectively, as the random processes $\bx(t)$ and $\by_j(t)$, for $t\geq 0$ and $j=1,2,\ldots,\bm_{\rm s}$. Here, $\bm_{\rm s}$ is the number of the sink nodes on a specific interval on the $x$ axis and is assumed to be a Poisson random variable with parameter $\lambda_s$ which is the average number of sink nodes per meter (i.e. density of the sink nodes on the $x$-axis). The coordinates of the nodes, based on the Wiener model, are given by
\begin{eqnarray}
\bx(t)&=&\sigma_0 \bnu_0(t)\label{brmodel0}\\
\by_j(t)&=&\by_j(0) + \sigma_j \bnu_j(t),\;\;j=1,2,\ldots,\bm_{\rm s}\label{brmodel}
\end{eqnarray}
where $\{\by_j(0)\}_{j=1}^{\bm_{\rm s}}$ are the random variables of the positions of the sink nodes at $t=0$, whereas $\sigma_0$ and $\sigma_j$ are the \emph{mobility parameters} of the sensor node and the $j$th sink node, respectively. The random processes $\{\bnu_j(t)\}_{j=0}^{\bm_{\rm s}}$  are statistically independent standard Wiener processes. Denoting the time dependent probability density function of the  process $\bnu_j(t)$ at time $t$ as $f_{\bnu_j}(x;t)$ and the joint density function of the processes $\bnu_i(t)$ and $\bnu_j(t)$ at $t_1$ and $t_2$ as $f_{\bnu_i,\bnu_j}(x_1,x_2;t_1,t_2)$, we can write
\begin{eqnarray}
  \bnu_j(t)\sim{\cal N} (0,t),\; \bnu_j(0)=0,\;\;j=0,1,\ldots,\bm_{\rm s}\nn\\
 \displaystyle f_{\bnu_i,\bnu_j}(x_1,x_2;t_1,t_2)=f_{\bnu_i}(x_1;t_1)f_{\bnu_j}(x_2;t_2),\; \mbox{ for } i\neq j\nn.
  \end{eqnarray}
Here, we have used $\sim$ to show the distribution of a random variable and ${\cal N} (a,b)$ refers to a Gaussian distribution with mean $a$ and variance $b$. Based on (\ref{brmodel0}), at $t=0$, the sensor node is at the origin. {The mobility parameter $\sigma_0$ of the sensor node (similar to the parameters $\sigma_j$ for the sink nodes) characterizes the probabilistic level of mobility for the sensor node. The position of a sensor node with mobility parameter equal to 1 at time $t$ is a Gaussian centered at the origin (assuming that the sensor starts at the origin) and variance of $t$, while for a sensor node with mobility parameter of 2, the position at time $t$ is a Gaussian with variance $4t$. This indicated that among two sensors with different mobility parameters, the sensor node with a higher mobility parameter has a higher chance to get further from its initial point after an equal amount of time. Therefore, we use this parameter to control and quantify the mobility of the network nodes.}

 As the network evolves over time, the sensor node chooses $n$ time instants $\{t_k\}_{k=1}^{n}$, {$0<t_1<t_n<T$ }to transmit its data. %

%
\label{Markovian} It is worth mentioning that the Wiener process is Markovian, i.e., if the location of a node which is moving based on this model, is known at $t=t_0$, its location at any time $t>t_0$ is \emph{independent} of its location at any time $t<t_0$. Furthermore, if at $t=t_1$, a node is at location $x$, its location at $t=t_2>t_1$ is also a Wiener process and its distribution is ${\cal N}\left(x,\sigma^2(t_2-t_1)\right)$.

In the subsequent sections, we define three parameters to study the effects of the node mobility on the performance of the considered wireless sensor network. Assume that the sensor aims to transmit its data to at least one of the sink nodes in $n$ time instants $\{t_k\}_{k=1}^n$. In each time instant, if the sensor is in the $d$-proximity of at least one sink node, there is a chance for a successful communication between the sensor node and that sink node. Let us refer to  such a time instant as a \emph{covered time instant} meaning that the sensor is in the coverage range of at least one sink node at that time instant.

The first parameter to be studied is the average number of covered time instants within the $n$ time instants. As discussed in the previous section, depending on the adopted scenario, this parameter directly corresponds to the reliability gain or the multiplexing gain of the communication between node and the fusion center. The second parameter is the ``outage probability'' defined as the probability that the sensor is \emph{not} in the communication range of any of the sink nodes in any of the $n$ time instants. Alternatively, outage probability can be thought of as the probability that the number of covered time instants is zero. Finally, the third parameter that we consider here is the largest number of consecutive uncovered time instants. This parameter also represents the longest time interval between two consecutive sensor-to-sink communication opportunities. If the quantity that is being sensed, is time-sensitive, this parameter can be used to quantify the worst case scenario for the delay of communication from the sensor node to the fusion center.

\section{Average Number of Covered Time instants}
\label{ENsection}
The first performance parameter to be studied is the average number of covered time instants. This parameter measures the number of time instants, out of $n$ time instants, that the sensor node \emph{sees} at least one sink node in its communication range, and can therefore, transmit its data. We first derive an equation for this parameter and then prove that it increases when the mobility parameter of the sensor node $\sigma_0$ is increasing.

Let us first define the indicator random variable $\bI_k$ as
\begin{eqnarray}
  \bI_k=\left\{\begin{array}{cl}
  1,& \mbox{if the sensor node \emph{is} in the communication range of at least one sink node at } t=t_k.\\
  0,  & \mbox{if the sensor node \emph{is not} in the communication range of any sink node at } t=t_k.
  \end{array}\right.\label{Ik}
\end{eqnarray}
 The realizations of the random sequence $\{\bI_k\}_{k=1}^n$ with relatively high number of ones correspond to a relatively high number of communication opportunities within $n$ time instants. The random variable $\bN$ is defined as the number of ones in the random binary sequence $\{\bI_k\}_{k=1}^n$ or simply, $\bN=\sum_{k=1}^n\bI_k$. Therefore, $\bN$ denotes the number of time instants in which the sensor is potentially able to communicate its data to a sink node. We define our first communication performance parameter as the expectation of $\bN$ denoted by $\EN$.
As discussed in the previous section, the parameter $\bN$ is a random variable representing the number of independent channels (or number of independent topologies) that the sensor can use to transmit its data while aiming for diversity gain and/or multiplexing gain. As $\bN$ is a random variable that will vary with the realization of the node time-varying location processes, $\EN$ is adopted to measure the performance of the communication scheme. Relatively larger values of $\EN$ translate into higher diversity and/or multiplexing gains.

   To investigate the effect of mobility on  $\mathbb{E}\{\bN\}$, we consider a scenario in which the sink nodes are distributed uniformly on {a segment of} the $x$-axis denoted by $[-D,\;D]$ with an average of $\lambda_s$ nodes per meter at $t=0$. {We further assume that based on the time interval $T$ and the mobility parameter of the sensor node $\sigma_0$, the choice of $D$  guaranties that the sensor node stays in this segment of the $x$-axis with probability one. In other words, we assume that within the $T$ seconds that the sensor attempts communicating with the sink nodes, it will \emph{not} leave the area covered by the sink nodes. This condition can be enforced by assuming that
   \begin{equation}
     \sigma_0^2 T\ll D \label{condotion}.
   \end{equation}  }

   {Based on the communication scheme adopted by the sensor node, if at the sensing time ($t=0$) a sink node is in the $d$-proximity of the sensor node, the sensed data will be transmitted to the server. If at $t=0$, the sensor is not covered by a sink node, then the communication will be  repeated $n$ times on the $\{t_k\}_{k=1}^{n}$, $0<t_1<t_n<T$ instants. To mathematically model the case where the sensor is not in the coverage of a sink node at $t=0$, we define the interval $J_D^{-h}\triangleq[-D,\;-h]\cup[h,\;D]$ and assume that the sink nodes are uniformly distributed on $J_D^{-h}$ at $t=0$, where $d<h\ll D$ is chosen to ensure that a sensor node at the origin is \emph{not} in the communication range of a sink node at the sensing time instant.}
    Also, the assumption of randomly distributed sink nodes with an average of $\lambda_s$ nodes per meter on the $x$-axis at $t=0$, necessitates that the number of sink nodes within $\jdh$, denoted by $\bm_{\rm s}$, be a Poisson random variable with mean $2D_h\times\lambda_s $, where $D_h\triangleq D-h$. That is, 
\begin{equation}\label{ms}
   {\rm Pr} \{\bm_{\rm s} = m \}=\displaystyle
  \frac{e^{-2\lambda_{\rm s} D_h} (2\lambda_{\rm s} D_h)^m}{m!},\; m=0,1,2,\ldots.
\end{equation}
We then assume that the sensor node will be communicating with one of these $\bm_s$ nodes.
The network nodes then start to move independently according to the Wiener process models described in (\ref{brmodel0}) and (\ref{brmodel}). We further assume that the sink nodes all share the same mobility parameter which is different from the mobility parameter of the sensor node. Mathematically,
this means\begin{equation}\label{samesigma}
  \sigma_j=\sigma\neq\sigma_0\;\;\mbox{ for } j=1,2,\ldots, \bm_{\rm s}.  \end{equation}
Based on the definition of $\bN$, we have
\begin{eqnarray}\label{EN0}
\mathbb{E}\{\bN\}=\sum_{k=1}^{n}{\rm Pr}\{\bI_k=1\}
\end{eqnarray}
where the expectation is taken with respect to the random locations of the nodes as well as the number of the sink nodes and their random initial locations. To calculate ${\rm Pr}\{\bI_k=1\}$, we use the Bayes rule to write
\begin{equation}\label{Bayes}
  {\rm Pr}\{\bI_k=1\}=\sum_{m=1}^{\infty}\pr\{\bI_k=1|\bm_{\rm s}=m\}\Pr\{\bm_{\rm s}=m\}.\\
\end{equation}
Note that $\pr\{\bI_k=1|\bm_{\rm s}=0\}=0$ and therefore, the summation in (\ref{Bayes}) starts from $m=1$.
Let us define the sequence of random processes $\bz_j(t)\triangleq |\bx(t)-\by_j(t)|$, as the distance between the
sensor node and the $j$th sink node at time $t$. At any time $t$, the sensor node has the possibility to transmit its data if at least one sink node is in its communication range or equivalently, if $\displaystyle\min_j \bz_j(t)\leq d$. Using this observation, and assuming that   at $t=0$, $m$ sink nodes are located in the interval $\jdh$ along the $x$-axis, we can write
\begin{eqnarray}
\pr\{\bI_k=1|\bm_s=m\}&=&\pr\{\min_j \bz_j(t_k)\leq d |\,\bm_s=m\}\nn\\
&=&1-\pr\{\min_j \bz_j(t_k)>d |\, \bm_s=m\}\nn\\
&=&1-\pr\{\bz_1(t_k)>d ,\ldots \bz_{m}(t_k)>d \}\label{zj}.
\end{eqnarray}
The random processes $\{\bz_j(t)\}_{j=1}^{\ms}$ are
not independent for different sink node index $j$, but it can be easily seen that they are independent
conditioned on the value of $\bx(t)$. Indeed, if $\bx(t)$ is
known, the processes $\{\bz_j(t)\}_{j=1}^{\ms}$ are only functions of
$\{\bnu_j(t)\}_{j=1}^{\ms}$, which are assumed to be
independent Wiener processes.
We hence use the Bayes rule to write
\begin{eqnarray}
  \pr\{\bz_1(t_k)>d ,\ldots, \bz_{m}(t_k)>d \}&=&
  \int_{x\in\mathbb{R}}{\pr\{\bz_1(t_k)>d ,\;\ldots,\; \bz_{m}(t_k)>d \;|\;\bx(t_k)=x\}f_{\bx}(x;t_k)dx}\nn\\
  &=&\int_{\mathbb{R}}{\prod_{j=1}^{m}\pr\{\bz_j(t_k)>d\;|\;\bx(t_k)=x\}}f_{\bx}(x;t_k)dx\label{zjcondx}
\end{eqnarray}
 where $f_{\bx}(x; t_k)$ is the probability density function of the coordinate of the sensor node at time $t=t_k$. Substituting (\ref{zjcondx}) into (\ref{zj}), we get
\begin{eqnarray}
\pr\{\bI_k=1|\bm_{\rm s}=m\}&=&1-\int_{x\in\mathbb{R}} \prod_{j=1}^{m}\pr\{\bz_j(t_k)>d\, \;| \ \bx(t_k)=x\} f_{\bx}(x; t_k)dx\nn\\
&=&1-\int_{\mathbb{R}}\prod_{j=1}^{m}
\left(1-\pr\{x-d\leq\by_j(t_k)\leq x+d\}\right) f_{\bx}(x ; t_k)dx\nn\\
&=&1-\int_{\mathbb{R}} \prod_{j=1}^{m}
\left(\int_{ {\bar J_d}(x)}f_{\by_j}(y; t_k)dy\right)\ f_{\bx}(x;t_k)dx\label{EN2}
\end{eqnarray}
where $J_d(x)\triangleq[x-d,x+d]$, ${\bar J_d}(x)\triangleq\mathbb{R}\smallsetminus J_d(x)$, and using \eqref{brmodel0}, the probability density function of the coordinate of the sensor node at time $t=t_k$, denoted by $f_{\bx}(x; t_k)$,  can be written as \begin{equation}
  f_{\bx}(x; t_k)=\frac{1}{\sigma_0 \sqrt{2\pi  t_k}}
  e^{-\displaystyle\frac{x^2}{2\sigma_0^2 t_k}}.
\end{equation}
Let us denote
\begin{equation}\label{gj}
  g_j(x,t_k) \triangleq \int_{ {\bar J_d}(x)}f_{\by_j}(y;t_k)dy=1-\pr\{x-d\leq\by_j(t_k)\leq x+d\}
\end{equation}
which  is the probability that at time $t_k$, the $j$th sink node is not in the communication range of a sensor positioned at coordinate $x$.
Substituting (\ref{EN2}) into (\ref{Bayes}), $\Pr\{\bI_k=1\}$ can be written as
\begin{eqnarray}
  {\rm Pr}\{\bI_k=1\}&=&\sum_{m=1}^{\infty}\pr\{\bI_k=1|\bm_{\rm s}=m\}\Pr\{\bm_{\rm s}=m\}\nn\\
    &=&\displaystyle\sum_{m=1}^{\infty}\Pr\{\bm_{\rm s} = m\}-\displaystyle\sum_{m=1}^{\infty}\int_{{\mathbb{R}}} \prod_{j=1}^{m}g_j(x,t_k) {\rm Pr}\{\bm_{\rm s} = m\} f_{\bx }(x;t_k)dx\nn\\
&=&1-e^{-2\lambda_{\rm s}D_h}-\displaystyle\sum_{m=1}^{\infty}\int_{{\mathbb{R}}} g^m(x,t_k) {\rm Pr}\{\bm_{\rm s} = m\} f_{\bx }(x;t_k)dx\label{EN3}
\end{eqnarray}
where in the last equation, we have used the fact that $\sum_{m=1}^{\infty}\Pr\{\bm_{\rm s} = m\} = 1 -\Pr\{\bm_{\rm s} = 0\} = 1- e^{-2\lambda_s D_h}$   and the assumption of equal node mobility parameters for the sink nodes in (\ref{samesigma}). This assumption means that   $g_j(x,t_k)$ is the same for all $j$, hence we drop the subscript $j$, and hereafter, use $g(x,t_k)$ instead of   $g_j(x,t_k)$.

Substituting (\ref{EN3}) into (\ref{EN0}) yields
 \begin{align}\label{ENfinal}
 \mathbb{E}\{\bN\}=n\left(1-e^{-2\lambda_{\rm s}D_h}\right)-\sum_{k=1}^{n}\sum_{m=1}^{\infty}\int_{\mathbb{R}}g^m(x,t_k){\rm Pr}\{\bm_{\rm s}=m\}f_{\bx}(x;t_k)dx.
\end{align}
 Using the distribution of $\bm_s$ given in (\ref{ms}), we can further simplify (\ref{ENfinal}) as
 \begin{eqnarray}
    \mathbb{E}\{\bN\}&=&n\left(1-e^{-2\lambda_{\rm s}D_h}\right)-\sum_{k=1}^{n}\sum_{m=1}^{\infty}\int_{\mathbb{R}}g^m(x,t_k)e^{-2\lambda_s D_h}\displaystyle\frac{ (2\lambda_s D_h)^m}{m!}f_{\bx}(x;t_k)dx\nn\\
&=&n\left(1-e^{-2\lambda_{\rm s}D_h}\right)-\sum_{k=1}^{n}\int_{\mathbb{R}}e^{-2\lambda_s D_h}\sum_{m=1}^{\infty}\displaystyle\frac{ (2g(x,t_k)\lambda_s D_h)^m}{m!}f_{\bx}(x;t_k)dx\nn\\
&=&n\left(1-e^{-2\lambda_{\rm s}D_h}\right)-\sum_{k=1}^{n}\int_{\mathbb{R}}\displaystyle e^{-2\lambda_s D_h}\left(e^{2g(x,t_k)\lambda_s D_h}-1\right)f_{\bx}(x;t_k)dx\nn
\nonumber \\  &=&n-\sum_{k=1}^n\int_{\mathbb{R}}f_{\bx}(x;t_k)e^{-2\lambda_s D_h(1-g(x,t_k))}dx\label{simplified}\,.
 \end{eqnarray}
{Please note that although the integration in \eqref{simplified} is on $\mathbb{R}$, based on \eqref{condotion}, the first term $f_{\bx}(x;t_k)$ will be non-zero only for values of $x$ for which $|x|\ll D$. This implies that the behaviour of the term after $f_{\bx}(x;t_k)$ is of our interest only for those values of $x$.}
In the next section, we will analyze the result in (\ref{simplified}) further and show that $\EN$ is an increasing function of the mobility parameter of the sensor node, $\sigma_0$. The function $g(x,t_k)$ plays an important role in studying the behavior of $\EN$.  Using (\ref{gj}), $g(x,t_k)$ can be further expanded as
\begin{eqnarray}
  g(x,t_k)&=&1-\pr\{x-d\leq\by_j(t_k)\leq x+d\}\nn\\
  &=&1-\int_{y\in\jdh}\pr\{x-d\leq\by_j(t_k)\leq x+d\;|\; \by_j(0)=y\}f_{\by_j}(y;0)dy\nn\\
  &=&1-\int_{y\in\jdh}\int_{y'=x-d}^{x+d}f_{\by_j(t_k)|\by_j(0)}(y'\;|\;y)f_{\by_j}(y;0)dy' dy \label{gj1}
\end{eqnarray}
where, in order to remove the effect of indexing of the sink nodes, we have assumed that the sink nodes are numbered randomly, so that the $j$th sink node can be anywhere on the segment $\jdh$ at $t=0$. This yields $f_{\by_j}(y;0)=1/(2D_h)$ for $y\in\jdh$. Also, $f_{\by_j(t_k)|\by_j(0)}(y'|y)$ is the conditional probability density function of the process $\by_j(t)$ at $t=t_k$ provided that $\by_j(0)=y$.  The Markovian property of the Wiener processes $\by_j(t)$ implies that this conditional pdf can be written as
\begin{eqnarray}
f_{\by_j(t_k)|\by_j(0)}(y'\;|\;y)={\cal N}(y,\sigma^2 t_k)=\frac{1}{\sigma\sqrt{2\pi t_k}}\exp\{-\frac{(y'-y)^2}{2\sigma^2 t_k}\}\nn.
\end{eqnarray}
Using this assumption, we can simplify (\ref{gj1}) as
\begin{eqnarray}
  g(x,t_k)&=&1-\frac{1}{2D_h\sigma\sqrt{2\pi t_k}}
  \left\{\int_{\jdh}\int_{x-d}^{x+d}\exp\{-\frac{(y'-y)^2}{2\sigma^2 t_k}\}dy' dy\right\} \label{17new}\\
  &=&1-\frac{1}{2 D_{h}}\int_{\jdh}\left\{Q\left(\displaystyle\frac{x-d-y}{\sigma\sqrt{t_k}}\right)
  -Q\left(\displaystyle\frac{x+d-y}{\sigma\sqrt{t_k}}\right)\right\}dy.\label{gk}
  \end{eqnarray}
Here, we define $\eta \triangleq 2D_h\sigma\sqrt{2\pi t_k}$ and $Q(x)\triangleq\frac{1}{\sqrt{2\pi}}\int_x^\infty e^{-\tau^2/2}d\tau$.

\section{Mobility Increases the Average Number of Covered Time instants}

 We now rigorously investigate the effect of increasing the mobility parameter of the sensor node $ \sigma_0$ on the average number of covered time instants denoted by $\EN$ which was derived earlier. We prove that $\EN$ as derived in (\ref{simplified}), is an increasing function of the sensor mobility parameter $\sigma_0$.

To do so, we consider a scenario where the sensor node is not in the communication range of any of the sink nodes at $t=0$. We then prove that the chance of being in a communication range of a sink node at any of the next time instants, $\Pr\{\bI_k=1\}$ for $ k=1,2,\ldots,n\;$ is an increasing function of the sensor mobility parameter. To realize this scenario, the parameter $d<h\ll D$ is used to ensure that the sensor node positioned at the origin at $t=0$, is \emph{not} in the communication range of any sink node at $t=0$.

 We first study the function $g(x,t_k),$ given as in (\ref{gk}) . Through the following lemmas, we prove that this function has a single maximum at $x=0$ and is monotonically increasing for negative values of $x$ and monotonically decreasing for positive values of $x$.\newline
   \emph{\textbf{Lemma 1:}} \emph{The function $p(y,h,\sigma^2)=\displaystyle\int_{x=-h}^h{e^{-\displaystyle\frac{(x-y)^2}{2\sigma^2}}}dx$ is a decreasing function of $|y|$. }\newline
    \emph{\textbf{Proof:}} Differentiating $p(y,h,\sigma^2)$ with respect to $y$ yields:
    \begin{eqnarray}
      \frac{d     p(y)}{dy}=\int_{-h}^h{2\frac{(x-y)}{2\sigma^2}e^{-\frac{(x-y)^2}{2\sigma^2}}}dx=-(e^{-\frac{(h-y)^2}{2\sigma^2}}-e^{-\frac{(h+y)^2}{2\sigma^2}})=-2\displaystyle e^{-\frac{(h^2+y^2)}{2\sigma^2}}\sinh{(\frac{2hy}{2\sigma^2})} \label{lemma1}
    \end{eqnarray}
    which is positive for $y<0$ and negative for $y>0$. Therefore, if $0<y_2<y_1$, we have $p(y_1,h,\sigma^2)<p(y_2,h,\sigma^2),$ and if $y_1<y_2<0$, we have $p(y_1,h,\sigma^2)<p(y_2,h,\sigma^2).  $ The proof is complter

\hfill $\blacksquare$    

Lemma 1 implies that if $|y_1|<|y_2|$ then $p(y_1,h,\sigma^2)>p(y_2,h,\sigma^2)$.

  \emph{\textbf{Lemma 2:}} \emph{The function $g(x,t_k)$ has the following properties}
  \begin{eqnarray}
      \left\{\begin{array}{cc}\displaystyle\frac{\partial g(x,t_k)}{\partial x}>0 & \mbox{ for } x<0\\ \displaystyle\frac{\partial g(x,t_k)}{\partial x}=0 & \mbox{ for } x=0 \\ \displaystyle\frac{\partial g(x,t_k)}{\partial x}<0 & \mbox{ for }x >0\end{array}\right.
  \end{eqnarray} {for $|x|\ll D$}.

\begin{figure}
\centering
\begin{tabular}{cc}
\psfrag{for x=0}{for $x=0$}
\psfrag{for x>0}{for $x>0$}
\psfrag{-h}{\small $\;-h$}
\psfrag{h}{\small $h$}
\psfrag{-d}{\small $\;-d$}
\psfrag{d}{\small $d$}
  \includegraphics[width=85mm]{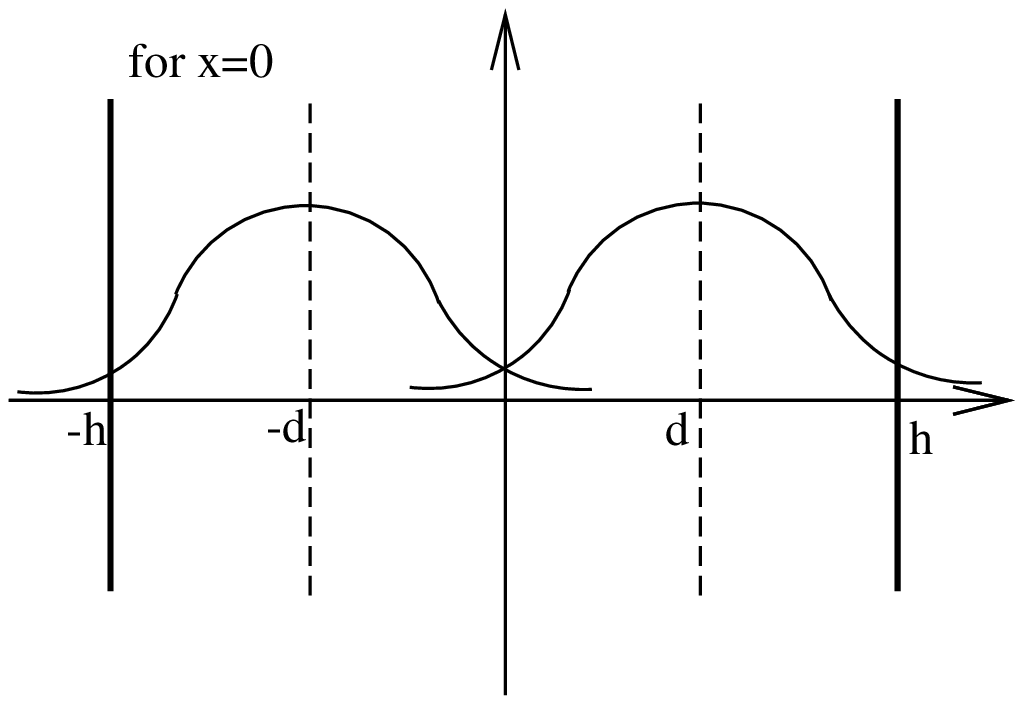}&
\psfrag{for x>0}{for $x>0$}
\psfrag{-h}{\small $\;\;\;-h$}
\psfrag{h}{\small$h$}
\psfrag{x-d}{\small$x-d$}
\psfrag{x+d}{\small$x+d$}
  \includegraphics[width=85mm]{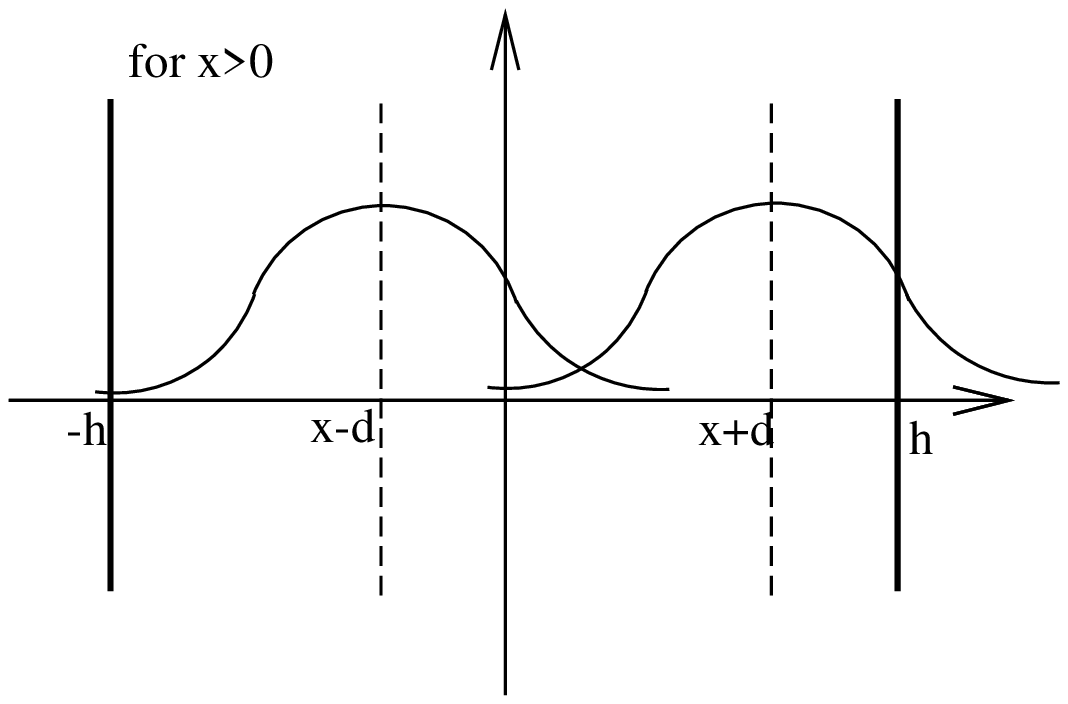}\\
  I&II
\end{tabular}
\caption{Derivative of $g(x;t_k)$ involves subtractions of the areas of two Gaussians centered at $x+d$ and $x-d$ from $-h$ to $h$. In the first figure above, $x=0$ and the Gaussians have similar area within the range. For positive $x$, the Gaussian centered at $x+d$ has always smaller area within $(-h,h)$ and hence the subtraction is negative.}\label{integrals}
\end{figure}

    \emph{\textbf{Proof:}} Differentiating $g(x,t_k)$ in (\ref{17new}) with respect to $x$  yields
    \begin{eqnarray}
\frac{\partial g(x,t_k)}{\partial x}=\frac{-1}{\eta}\int_{y\in J_D^{-h}} \left(\exp\left\{-\frac{(y-(x+d))^2}{2\sigma_j^2 t_k}\right\}dy-  \exp\left\{-\frac{(y-(x-d))^2}{2\sigma_j^2 t_k}\right\}\right)dy\\=\frac{1}{\eta}\overbrace{\left\{\int_{y=-h}^h \exp\left\{-\frac{(y-(x+d))^2}{2\sigma_j^2 t_k}\right\}dy-  \int_{y=-h}^h \exp  \left\{-\frac{(y-(x-d))^2}{2\sigma_j^2 t_k}\right\}dy\right\}}^{A}-\nonumber \\\;&\;\\
\frac{1}{\eta}\underbrace{\left\{\int_{y=-D}^D \exp\left\{-\frac{(y-(x+d))^2}{2\sigma_j^2 t_k}\right\}dy-  \int_{y=-D}^D \exp\left\{-\frac{(y-(x-d))^2}{2\sigma_j^2 t_k}\right\}dy\right\}}_{B}.
\end{eqnarray}
For $x=0$, this derivative is zero because the two Gaussian distributions in braces $A$ and $B$ will be centered at $d$ and $-d$ (Fig.~\ref{integrals}-I) and hence will have similar areas within $(-h,h)$ (for the first brace) and $(-D,D)$ (for the second brace). Therefore, the difference of the two Gaussian functions  in each brace will be zero and the total will also be zero.

 Let us now study the difference of two Gaussian functions for $x\neq 0$. Using the function $p(y,h,\sigma^2)$ defined in Lemma-1, we can write
 \begin{equation}
   A=p(x+d,h,\sigma^2 t_k)-p(x-d,h,\sigma^2 t_k).
 \end{equation}
 As $d>0$, for positive values of $x$, we know that $|x+d|>|x-d|$ and therefore, using Lemma 1, $p(x+d,h,\sigma^2 t_k)<p(x-d,h,\sigma^2 t_k)$ and $A<0$. For negative values of $x$,  $|x+d|<|x-d|$ and Lemma 1 leads to $A>0$. To summarize, we have justified that
\begin{eqnarray}
A= \int_{y=-h}^h \exp\left\{-\frac{(y-(x+d))^2}{2\sigma_j^2 t_k}\right\}dy-  \int_{y=-h}^h \exp\left\{-\frac{(y-(x-d))^2}{2\sigma_j^2 t_k}\right\}dy \mbox{ is }\left\{\begin{array}{cc}\mbox{positive} & x<0\\ 0 & x=0 \\ \mbox{negative}  & x>0\end{array}\right..\nn
\end{eqnarray}

{Based on the assumption of $|x|\ll D$, the two Gaussians in the second brace will have negligible areas outside the $(-D,D)$ boundary and therefore the values of the two integrals in $B$ are both 1 and the second brace is zero ($B=0$). }
This implies that
\begin{equation}
  \frac{\partial g(x,t_k)}{\partial x}<0 \;\; \mbox{for  } 0<x\ll D\;\;\
  \mbox{and}\;\;\frac{\partial g(x,t_k)}{\partial x}>0 \;\; \mbox{for  } -D\ll x<0\nn
\end{equation}
meaning that $g(x,t_k)$ has a unique maximum at $x=0$ for $|x|\ll D$. The proof is complete.\hfill $\blacksquare$
\newline \emph{\textbf{Lemma 3:} The function $\ell(x)=\exp(-2\lambda_s D_h(1-g(x,t_k)))$ has the property that $x \displaystyle\frac{\partial \ell(x)}{\partial x} \le 0$ for { $0<|x|\ll D$}}.\newline
\emph{\textbf{Proof}}: Using Lemma 2, we already know that the function $g(x,t_k)$ has a single maximum at $x=0$ {for $|x|\ll D$}. From (\ref{EN2}) and (\ref{EN3}) we know that $g(x,t_k)=\pr\left\{\bz_j(t_k)>d\, \Big| \, \bx(t_k)=x\right\}$ represents the probability of an event and hence is limited between $0$ and $1$. Therefore, $1-g(x,t_k)$ is a function with a minimum at $x=0$. This implies that $\ell(x)=\exp(-2\lambda_s D_h(1-g(x,t_k)))$, similar to
$g(x,t_k)$, has a maximum at $x = 0$, and has a positive derivative for negative $x$ and negative derivative for positive $x$. Hence, $x \displaystyle\frac{\partial \ell(x)}{\partial x}\le0$ for all {$0<x\ll D$} and this concludes the proof.\hfill $\blacksquare$
\newline It is worth mentioning that $\ell(x)$ is also a function of other model variables ($\lambda_s$, $d$, $D$, $t_k$ and $h$). However, in order to avoid notation complexity and as we are only interested in the functionality of $\ell$ in $x$, we dropped other parameters and just used the notation $\ell(x)$.

Now, we are well-positioned to prove the following theorem:

\emph{\textbf{Theorem 1}: $\mathbb{E}\{\bN\}$ is an increasing function of the mobility parameter of the sensor, $\sigma_0$.}\newline
\emph{\textbf{Proof}}:
Differentiating $\EN$  in (\ref{simplified}) with respect to the mobility parameter of the sensor node, $\sigma_0$, yields
\begin{eqnarray}
  \frac{\partial \EN}{\partial \sigma_0}&=&\frac{\partial}{\partial \sigma_0}\left(n-\sum_{k=1}^n{\int_{\mathbb{R}}{\frac{1}{\sigma_0\sqrt{2\pi t_k}}e^{-\frac{x^2}{2{\sigma_0}^2 t_k}}\ell(x)dx}}\right)\nn\\
&=&-\sum_{k=1}^n{\frac{\partial}{\partial \sigma_0} \int_{\mathbb{R}}{\frac{1}{\sigma_0\sqrt{2\pi t_k}}e^{-\frac{x^2}{2{\sigma_0}^2 t_k}}\ell(x)dx}}\nn\\
&=&\sum_{k=1}^n \left( \int_{\mathbb{R}}{\frac{1}{\sigma_0^2\sqrt{2\pi t_k}}e^{-\frac{x^2}{2{\sigma_0}^2 t_k}}\ell(x)dx} \right) -\int_{\mathbb{R}}\frac{x^2}{{\sigma_0}^4t_k\sqrt{2\pi t_k}}e^{-\frac{x^2}{2\sigma_0^2 t_k}}\ell(x)dx\label{intby1}\\
&=&\sum_{k=1}^n { \int_{\mathbb{R}}{\frac{1}{\sigma_0^2\sqrt{2\pi t_k}}e^{-\frac{x^2}{2{\sigma_0}^2 t_k}}\ell(x)dx} }-\nn\\
&&\sum_{k=1}^n\left(\frac{1}{\sigma_0^2\sqrt{2\pi t_k}}\left[x\ell(x)e^{-\frac{x^2}{2{\sigma_0}^2 t_k}}\right]_{-\infty}^{\infty}+\frac{1}{\sigma_0^2\sqrt{2\pi t_k}}\int_{\mathbb{R}}{e^{-\frac{x^2}{2\sigma_0^2 t_k}}}\left[\ell(x)+xd\ell'(x)\right]dx\right)\label{intby2}\\
&=&-\sum_{k=1}^n\frac{1}{\sigma_0^2\sqrt{2\pi t_k}}\int_{\mathbb{R}}{x\ell'(x) e^{-\frac{x^2}{2\sigma_0^2 t_k}}dx} \label{proof3}.
\end{eqnarray}
Here, $\ell'(x)=\partial \ell(x)/\partial x$ and integration-by-part has been used to expand (\ref{intby1}) to (\ref{intby2}).
 Based on Lemma 2,  $x\partial \ell(x)/\partial x$ is negative for {all $|x|\ll D$, and therefore, in the last integral in (\ref{proof3}), the integrand is negative for $|x|\ll D$ and zero otherwise. The integral, hence, has a negative value for all $k$}. This means that $\displaystyle\frac{\partial \EN}{\partial \sigma_0}>0$ and the proof is complete.\hfill $\blacksquare$

It is now analytically shown that increasing the mobility of a sensor node increases its chances of being exposed to a sink node in the future if it is not currently in the communication range of one. Consequently, if we are observing the sensor node over a number of time instants, the average number of instants in which the sensor is exposed to a sink node also increases with increasing mobility. The rate of increase in $\EN$ however slows down when the node mobility parameter grows large and saturates as discussed in the last paragraph of Section~\ref{ENsection}.

\section{Outage Probability and the Longest Transmission Delay}
\subsection{Outage Probability} Another important performance parameter is the probability of communication outage. In this section, we consider the same scenario considered the previous sections, where the sensor node is exploiting the change in the topology to obtain diversity and/or multiplexing gain by attempting to communicate with the sink nodes during $n$ time instants $\{t_k\}_{k=1}^n$. Outage happens when neither of these attempts is successful. In other words, if the sensor node is not in the communication range of any of the sink nodes during these $n$ time instants, an outage occurs. The probability of outage in this context is therefore defined as $\Pr\{\bN=0\}$.

We first express the outage probability in terms of the node mobility parameter and then prove that as the node mobility increases, the probability of outage decreases. To do so, we first use the Markovian property of the Wiener mobility model to calculate the probability of outage.

As mentioned earlier, the processes $\bx(t)$ and $\by_j(t)$ (coordinates of the sensor node and the sink nodes) are assumed to be Markovian. Therefore, for any $j$, the sequence $\{\bz_j(t_k)\}_{k=1}^n$, defined as the distance of the sensor node to the $j$th sink node at time $t_k$, will also be Markovian as this sequence is a function of two Markovian processes. Any function of these $\ms$ random processes will also form a Markovian process. Specifically, $\bz_{\rm min}(t_{k})=\min_j\{\bz_j(t_k)\}$ has the Markovian property.
Recalling that $\bI_k$ is the indicator random variable for the event $\min_j\{\bz_j(t_k)\}<d$, one can easily conclude that the Markovian property is inherited by the sequence $\bI_k$ as well. In fact, knowing the positions of the sink nodes at any time $t_k$ determines $\bI_k$ and also makes the position of the sensor nodes at any future time $t>t_k$, independent of their positions at $t<t_k$. As the positions of the sensor nodes will also result in a value of 0 or 1 for $\bI_k$, $\bI_{k+1}$ will hence be independent of $\bI_{k-1}$.

To calculate $\Pr\{\bN=0\}$, we first use the Bayes rule to to write this quantity as

\begin{eqnarray}
\Pr\{\bN=0\} &=&\Pr\{\bI_1=0,\bI_2=0,\ldots,\bI_n=0\}\nn\\
&=&\Pr\{\bI_n=0|\bI_1=0,\ldots,\bI_{n-1}=0\}\nn\\
&\times&\Pr\{\bI_{n-1}=0|\bI_1=0,\ldots,\bI_{n-2}=0\}\times\ldots\nn\\
&\times&\Pr\{\bI_2=0|\bI_1=0\}\Pr\{\bI_1=0\}\label{IkSequence}.
\end{eqnarray}
Now, we use the Markovian property of $\{\bI_k\}_{k=1}^n$ to further simplify (\ref{IkSequence}) as
\begin{eqnarray}
\Pr\{\bN=0\} &=&\Pr\{\bI_1=0\}\prod_{k=2}^{n}\Pr\{\bI_k=0|\bI_{k-1}=0\}\nn\\
&=&\Pr\{\bI_1=0\}\prod_{k=2}^{n}p_{00}^k.\label{outage}
\end{eqnarray}
Here, $p_{00}^k\triangleq\Pr\{\bI_k=0|\bI_{k-1}=0\}$ is defined as the probability that the sensor is not covered at time $t_k$, provided that it was not covered at time $t_{k-1}$ either. Similarly, we can define $p_{01}^k\triangleq\Pr\{\bI_k=1|\bI_{k-1}=0\}$ as the probability that the sensor node is covered at $t=t_k$, provided that it was not covered at $t=t_{k-1}$.

\emph{\textbf{Theorem 2}: The transition probability $p_{01}^{k}=\Pr\{\bI_k=1|\bI_{k-1}=0\}$ is an increasing function of the node mobility parameter $\sigma_0$ and the transition probability  $p_{00}^{k}=1-p_{01}^{k}$ is a decreasing function of $\sigma_0$}.\newline
\emph{\textbf{Proof}}: Note that the probability $\Pr\{\bI_k=1|\bI_{k-1}=0\}$ refers to the situation where the sensor node located at $\bx(t_{k-1})$ is \emph{not} covered by any of the sink nodes positioned at $\by_j(t_{k-1})$, while at $t=t_k$, at least one sink node exists within the $d$-proximity of the sensor node. Let us fix another axis, namely $x'$-axis with origin at $\bx(t_{k-1})$ and start another Wiener  process $\bx'(t')$ with time reference $t'=t-t_{k-1}$ so that
\begin{equation}
\bx'(t')=\bx'(t-t_{k-1})\triangleq  \bx(t)-\bx(t_{k-1})\;\;\mbox{ for } t\geq t_{k-1}\, .
\end{equation}
As the location of the sensor at $t=t_{k-1}$ is already observed to be $\bx(t_{k-1})$, the process $\bx'(t')$ is the same Wiener process (with the same mobility parameter) shifted in time and space so that at $t'=0$, it is located at the origin. A similar procedure can be adopted to define the Wiener processes of the sink node locations in the new time and space coordinates, $\by_j(t')$, for $j=1,2,\ldots\bm_s$, so that at $t'=0$, they are within the same relative distance from the sensor node as the $\by_j(t)$'s are at $t=t_{k-1}$. This is possible by choosing
\begin{equation}
  \by'_j(t')=\by'_j(t-t_{k-1})\triangleq\by_j(t)-\bx(t_{k-1})\;\;\;\mbox{ for } j=1,2,\ldots\bm_s.
\end{equation}
This ensures that the sensor node and the sink nodes in the shifted time and space coordinates maintain the same relative distance as the original processes. Based on this analogy, $\bI_{k-1}=0$ corresponds to $\bx'(t')$ \emph{not} being covered at $t'=0$ and $\bI_k = 1 $ corresponds to $\bx'(t')$ being in a $d$-proximity of at least one sink node at $t'=t_k-t_{k-1}$. Now let us define the indicator random variables $\bI'_k$ for the process $\bx'(t')$ similarly to the indicator random variable $\bI_k$ defined in (\ref{Ik}). With this definition, it is readily seen that
\begin{eqnarray}
p_{01}^{k}&=&\Pr\{\bI_k=1|\bI_{k-1}=0\}\nn\\
&=&\Pr\{\bI'_1=1\}\;\;\; \mbox{ for } t'_1=t_{k}-t_{k-1}
\end{eqnarray}
In other words, the $p^k_{01}$ in the $\bx(t)$ process, is equal to the $\Pr\{\bI'_1=1\}$ for the time- and space-shifted process $\bx'(t')$, if we choose the sequences $\{t'_k\}_{k=1}^n$ so that $t'_1 = t_k-t_{k-1}$. The Wiener random processes $\bx'(t')$ and $\by'_j(t')$ have all the properties required for for application of  Theorem 1, meaning that for each choice of the sequences $\{t'_k\}_{k=1}^n$ and for any choice of the time instants $t_k$, $\Pr\{\bI'_k=1\}$ is an increasing function of the mobility parameter of the process $\bx'(t')$. This also includes $\Pr\{\bI'_1=1\}$ for $t'_1=t_k-t_{k-1}$, and therefore, $p_{01}^{k}=\Pr\{\bI_k=1|\bI_{k-1}=0\}$ is an increasing function of the mobility parameter of the processes $\bx(t)$ and $\bx'(t')$ and $p_{00}^{k}=1-p_{01}^{k}$ is a decreasing function of the mobility parameter. \hfill $\blacksquare$
%
%

\emph{\textbf{Theorem 3}: The probability of outage (\ref{outage}) is a decreasing function of the sensor's mobility parameter,  $\sigma_0$.}\newline
\emph{\textbf{Proof}}: Based on Theorems 1 and Theorem 2, both $\Pr\{\bI_1=0\}$ and the sequence $\left\{p^{k}_{00}\right\}_{k=1}^n$ are decreasing functions of the mobility parameter $\sigma_0$, therefore, the probability of outage in (\ref{outage}) is also a decreasing function of the mobility parameter of the sensor node.\hfill $\blacksquare$

\subsection{Longest number of consecutive uncovered time instants}
While the first two discussed performance parameters, namely $\EN$ and $\Pr\{\bN=0\}$, measure the communication performance of the network in terms of the sensor node mobility parameter, the longest number of consecutive non-covered time instants, denoted by the random variable $\bN_d$, provides us with a measure of delay in the network. Remember that the sequence $\{\bI_k\}_{k=1}^n$ is an indicator random variable which takes the value 1, if and only if the sensor node is in the coverage of at least one sink node at time $t_k$, and 0 otherwise. Therefore, a run of zeros with length $L$ in this sequence of length $n$, represents a situation where the sensor has not been able to connect to \emph{any} sink node for $L$ consecutive time instants. If, for example, the time instants are chosen to be equally distanced with $t_i-t_{i-1}=\Delta t$ for $i=2,3,\ldots,n$, such run of zeros corresponds to a delay of $L\Delta t$ in two consecutive communication opportunities.
 %
%
The parameter $\bN_d$ is therefore defined to be the length of the \emph{longest} run of zeros in the random Markovian sequence  $\{\bI_k\}_{k=1}^n$. To derive $\bN_d$ in terms of the sensor node mobility parameter $\sigma_0$, we note that the sequence $\{\bI_k\}_{k=1}^n$ is a Markovian binary sequence. Therefore, we need to study the statistics of the longest run of zeros in a Markovian binary sequence. Let us define the sequence $\{\beta_k\}_{k=1}^n$ to be the length of the runs of zeros in the sequence $\{\bI_k\}_{k=1}^{n}$. In fact $\{\beta_k\}_{k=1}^n$ is counting the number of zeros in each run of zeros that occur in $\{\bI_k\}_{k=1}^n$. As an example for $n=10$, and one sample of $\bI_1\bI_2\ldots\bI_{10}= 1100100011$ the corresponding $\{\beta_k\}_{k=1}^n$ sequence will be $0012012300$. The sequence $\{\beta_k\}_{k=1}^n$ is therefore Markovian and its transition probabilities can be written as
\begin{eqnarray}
\pr\{\beta_k=x|\beta_{k-1}=x-1>0\}&=&\pr\{\bI_k=0|\bI_{k-1}=0\}= p_{00}^k\nn\\
\pr\{\beta_k=0|\beta_{k-1}>0\}&=&\pr\{\bI_k=1|\bI_{k-1}=0\}=p_{01}^k\nn\\
\pr\{\beta_k=1|\beta_{k-1}=0\}&=&\pr\{\bI_k=0|\bI_{k-1}=1\}=p_{10}^k\nn\\
\pr\{\beta_k=0|\beta_{k-1}=0\}&=&\pr\{\bI_k=1|\bI_{k-1}=1\}=p_{11}^k\label{transition}.
\end{eqnarray}
Here, $p_{mn}^k=\Pr\{\bI_k=n|\bI_{k-1}=m\},\;\;m,n\in\{0,1\}$ are the transition probabilities of the Markovian sequence $\bI_k$ as defined while deriving the expression  for outage probability in (\ref{outage}).
%
The transition probabilities $p_{mn}^k,\;\;m,n\in\{0,1\}$ can be written in terms of the conditional probability of the process $\bx(t)$. The reason is that due to the Markovian property of the Wiener process $\bx(t)$, given $\bx(t_{k-1})$,  $\bx(t_{k})$ is independent of $\bx(t_{k-2})$. Based on this observation, for instant, we can write
%
\begin{eqnarray}
  p_{01}^k=\int_{x_{k-1}\notin \Omega}\int_{x_k\in \Omega}f_{\bx_k|\bx_{k-1}}(x_k|x_{k-1})dx_k dx_{k-1}\nn.
\end{eqnarray}
 where the set $\Omega$ denotes the union of all length $2d$ line segments centered at the location of each sink node at time $t_k$ ($\by_j(t_k)$) within which a sensor node has the opportunity to communicate to the sink node and
 \begin{equation}
 f_{\bx_k|\bx_{k-1}}(x_k|x_{k-1})=\displaystyle\frac{1}{\sqrt{2\pi{\sigma_v^2}(t_{k}-t_{k-1})}} e^{\displaystyle-\frac{1}{(t_{k}-t_{k-1})\sigma_v^2}(x_{k}-x_{k-1})^2}\label{conditional}.
 \end{equation}
Knowing $\bx(t)$ at any arbitrary time $t_1$, the distribution of $\bx(t_2)$ is $\bx(t_2)\sim {\cal N}(\bx(t_1),\sigma_0^2(t_2-t_1))$.
  The transition probabilities of the Markovian sequence $\beta_k$ can hence be calculated based on the parameters of the mobility model of the sensor and sink nodes. Note that $\bN_d=\max_k \beta_k$. In other words, $\bN_d$ is the largest number that the counting sequence $\beta_k$ will experience while counting the length of runs of zero in $\bI_k$. Based on this observation, we can find the cumulative probability mass function of $\bN_d$ as follows.
\begin{eqnarray}
          {\rm Pr}\{\bN_d<L\}&=& \pr\{\max_j \beta_j<L\}
          ={\rm Pr}\{\beta_L<L , \beta_{L+1}<L,\ldots ,\beta_n<L\}\label{step1}\\
        &=&\sum_{i_L=0}^{L-1}\sum_{i_{L+1}=0}^{L-1}\ldots\sum_{i_n=0}^{L-1}{\rm Pr}\{\beta_L=i_L , \beta_{L+1}=i_{L+1},\ldots, \beta_n=i_n\}\label{step2}\\
        &=&\sum_{i_L=0}^{L-1}\sum_{i_{L+1}=0}^{L-1}\ldots\sum_{i_n=0}^{L-1}{\rm Pr}\{\beta_L=i_L\}\prod_{m=L+1}^{n}{\rm Pr}\{\beta_m=i_m|\beta_{m-1}=i_{m-1}\}\label{step3}.
      \end{eqnarray}
      Note that, as the sequence $\{\beta_k\}_{k=1}^n$ is counting the number of zeros in $\{\bI_k\}_{k=1}^n$, the first $\beta_k$ that can be greater than or equal to $L$ is $\beta_L$. This justifies the fact that no $\beta_k$ needs to be considered in (\ref{step1}) with $k<L$. In simplifying (\ref{step2}) into (\ref{step3}), the Markovian property of $\beta_k$ has been used. As can be seen from (\ref{step3}), the cumulative probability mass function of $\bN_d$ depends on the transition probabilities of $\beta_k$, represented based on the transition probabilities of $\bI_k$ in (\ref{transition}), as well as on the probability mass function of $\beta_L$ (appearing in (\ref{step3}) as $\pr\{\beta_L=i_L\}$). The probability mass function of $\beta_L$ has been studied and derived in \cite{Eryilmaz06} as
\begin{eqnarray}
  \pr\{\beta_n=k\}=\left\{\begin{array}{cc}
  a(n)& k=0\\
  p_{10}^{n-k+1} a(n-k)\prod_{i=1}^{k-1} p_{00}^{n-i+1} & k=1,2,\ldots,n-1\\
  \pr\{\bI_1=0\} \prod_{i=1}^{n-1} p_{00}^{n-i+1} & k = n \end{array}\right.
\end{eqnarray}
where $a(j)=(p_{11}^{(j)}-p_{01}^{(j)})a(j-1)+p_{01}^{(j)}\;\; \mbox{ for  }j\geq 2, \mbox{ and } a(1)=\pr\{\bI_1=1\}$. Note that $\pr\{\beta_n=0\}$ is the probability of outage as derived in (\ref{outage}). This concludes the derivation of the CMF of the longest transmission delay, $\bN_d$. As $\bN_d$ is a positive discrete random variable, the expected value of the longest delay can be obtained from (\ref{step3}) using ${\mathbb E}\{\bN_d\}=\sum_{L=0}^{\infty}{(1-{\rm Pr}\{\bN_d<L\})}\nn$. Our expectation is that ${\mathbb E}\{\bN_d\}$  decreases with increasing the sensor mobility parameter. In other words, mobility decreases the probability of large delays. In the numerical evaluation section, we investigate the behavior of $\ENd$ with respect to the mobility parameter of the sensor node using numerical evaluation of (\ref{step3}) and observe that higher mobility  of the sensor node results in \emph{smaller} average longest transmission delays. This in turn confirms that the mobility also enhances the performance of the network in terms of the data transmission delay.
\section{{Application in Vehicular Network}}\label{app}
{Consider a highway lane in which vehicles are moving in a direction with an average velocity of $\mu$ meters per second. The actual instantaneous speed of the vehicles in the lane varies around $\mu$ and the variations mainly depend on factors such as the driving style of the driver, road condition, etc. The position of a particular vehicle at time $t$ can therefore be modeled using a drifted Wiener process as
\begin{equation}
  \bx(t)=\mu t + \sigma \bnu(t)
\end{equation}
with $\bnu(t)$ defined similar to  \eqref{brmodel0}. Here, the parameter $\sigma$ controls the variations of the vehicle's speed around the average speed of $\mu$. Assume that vehicles in the highway are sensing a phenomenon (spectrum occupancy, road or weather condition, etc) and sharing their sensing with the neighboring nodes. Assuming that vehicles in a lane move with an equal average speed, the same Wiener process will be governing the processes of the distanced between the vehicles in a lane. This means that the same results can be derived to show that increasing the speed variations for an individual vehicle, increases its chances for communicating its information with the neighboring vehicles in the same lane.}

\section{Two dimensional motion}
Now consider the case where the nodes in a network are moving in a two-dimensional plane and confined in a $(-D,D)\times(-D,D)$ square portion of this plane at $t=0$. With two simplifying assumptions, we can use the results of the previous sections to find $\EN$. First, let us assume that the motion of the nodes in $x$-axis and $y$-axis are independent and both follow our Wiener process model (perhaps with different parameters). Second, let us assume that in order to establish a successful link between a sensor node and a sink node, both their $x$ and $y$ coordinates should be within a specific range of each other (a square coverage area rather than a circle). With the above assumptions, the probability that the sensor is in a coverage of at least one sink node at $t=t_k$, denoted by $\Pr\{\bI_k=1\}$ can be written as
\begin{align}
  &\pr\{\bI_k=1\}=  \pr\{\bI^x_k=1\}  \pr\{\bI^y_k=1\}
\end{align}
where $\bI^x_k$ and $\bI^y_k$ are the equivalent one-dimensional coverage indicators in $x$ and $y$ direction, respectively. In other words, $\pr\{\bI^x_k=1\}$ is the probability that at time $t_k$, the sensor node's $x$ coordinate lies within distance $d$ of the $x$ coordinate of at least one sink node.

In the case where the mobility parameters in both directions are equal, we can write
\begin{align}
  &\pr\{\bI_k=1\}=  \left(1-\int_{x}f_{\bx}(x;t_k)e^{-2\lambda_s D_h(1-g(x,t_k))}dx\right)^2\label{2D}.
\end{align}
This implies that $\pr\{\bI_k=1\}$ will still be an increasing function of the mobility parameter $\sigma_0$. The same approach used for the one dimensional case can hence be used to prove that $\EN$ is an increasing function, and probability of outage is a decreasing function of the mobility parameter of the sensor node $\sigma_0$ in the two dimensional scenario.

Let us now relax the assumption that each sink node covers a square area around it instead of a circle. A simple technique can be used to find an upper bound and a lower bound for the case where the motions in $x$ and $y$ axes are still independent but the coverage around a sink node is a circle of radius $d$.

\begin{figure}
\begin{center}
  \includegraphics[width=80mm]{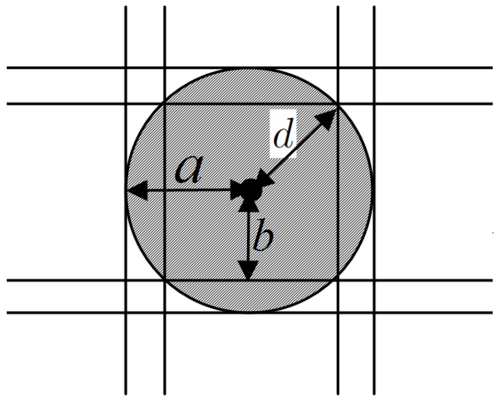}\\
  \caption{A simple method to bound the probability of successful transmission in 2-dimensional mobility model. Here $a=d$, and $b=d/\sqrt{2}$}\label{upperlower}
\end{center}
\end{figure}

As seen in Fig.~\ref{upperlower}, the probability of a sensor node being in $d$-proximity of a sink node at $t=t_k$ denoted by $\pr\{\bI_k=1\}$ (the shaded circle in Fig.~\ref{upperlower}), is larger than the probability of the sensor node being in the square inscribed by the circle (the smaller square) and smaller than the probability of the sensor node being in the square that inscribes the circle (the larger square). Therefore, the exact probability of a successful transmission is upper bounded by (\ref{2D}) with the range being the same $d$ and lower bounded by (\ref{2D}) with the range being $d/\sqrt{2}$.

As both the lower and upper bounds of $\pr\{\bI_k=1\}$ are increasing functions of the mobility parameter, one can deduce that mobility also enhances the performance of the sensor network in two dimensional scenarios.

\section{Numerical Evaluation}\label{numerical}
In this section, we use numerical evaluations to gain further insight into the behavior of the performance parameters introduced and the affect of mobility of the nodes on each parameter. {All the numerical evaluations presented here are obtained using MATLAB simulations with parameters described in each simulation.}

\begin{figure}
\centering
\begin{tabular}{cc}
\psfrag{lambdas}{\tiny $\lambda_s$}
\psfrag{EN}{\tiny $\mathbb{E}\{\bN\}$}
\psfrag{( sigmav =0.5)}{\small $\sigma_0=.5$}
\psfrag{ sigma0 =0.5}{\tiny $\sigma_0$ = 0.5}
\psfrag{ sigma0 =1}{\tiny $\sigma_0$ = 1}
\psfrag{ sigma0 =2}{\tiny $\sigma_0$ = 2}
\psfrag{ sigma0 =5}{\tiny $\sigma_0$ = 5}
\psfrag{ sigma0 =10}{\tiny $\sigma_0$ = 10}
\psfrag{ sigma0 =15}{\tiny $\sigma_0$ = 15}
\psfrag{ sigma0 =25}{\tiny $\sigma_0$ = 25}

  \includegraphics[width=85mm]{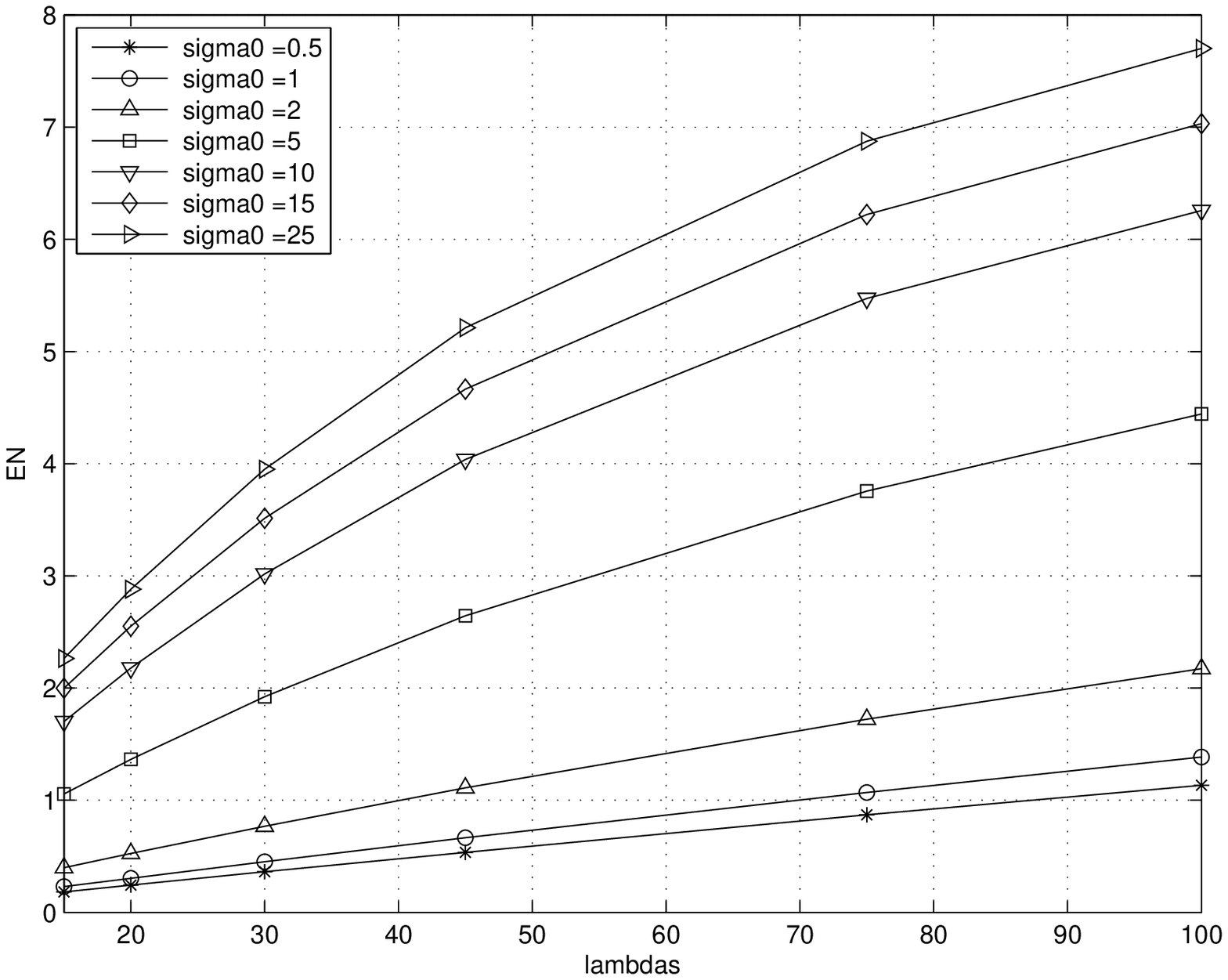}&
\psfrag{sigmav}{\tiny $\sigma_0$}
\psfrag{lambdas}{\tiny $\lambda_s$ Car per KM}
\psfrag{E{N}}{\tiny $\mathbb{E}\{\bN\}$}
\psfrag{ lambdas = 15}{\tiny$\lambda_s$ = 15}
\psfrag{ lambdas = 20}{\tiny$\lambda_s$ = 20}
\psfrag{ lambdas = 30}{\tiny$\lambda_s$ = 30}
\psfrag{ lambdas = 45}{\tiny$\lambda_s$ = 45}
\psfrag{ lambdas = 75}{\tiny$\lambda_s$ = 75}
\psfrag{ lambdas = 100}{\tiny$\lambda_s$ = 100}

  \includegraphics[width=85mm]{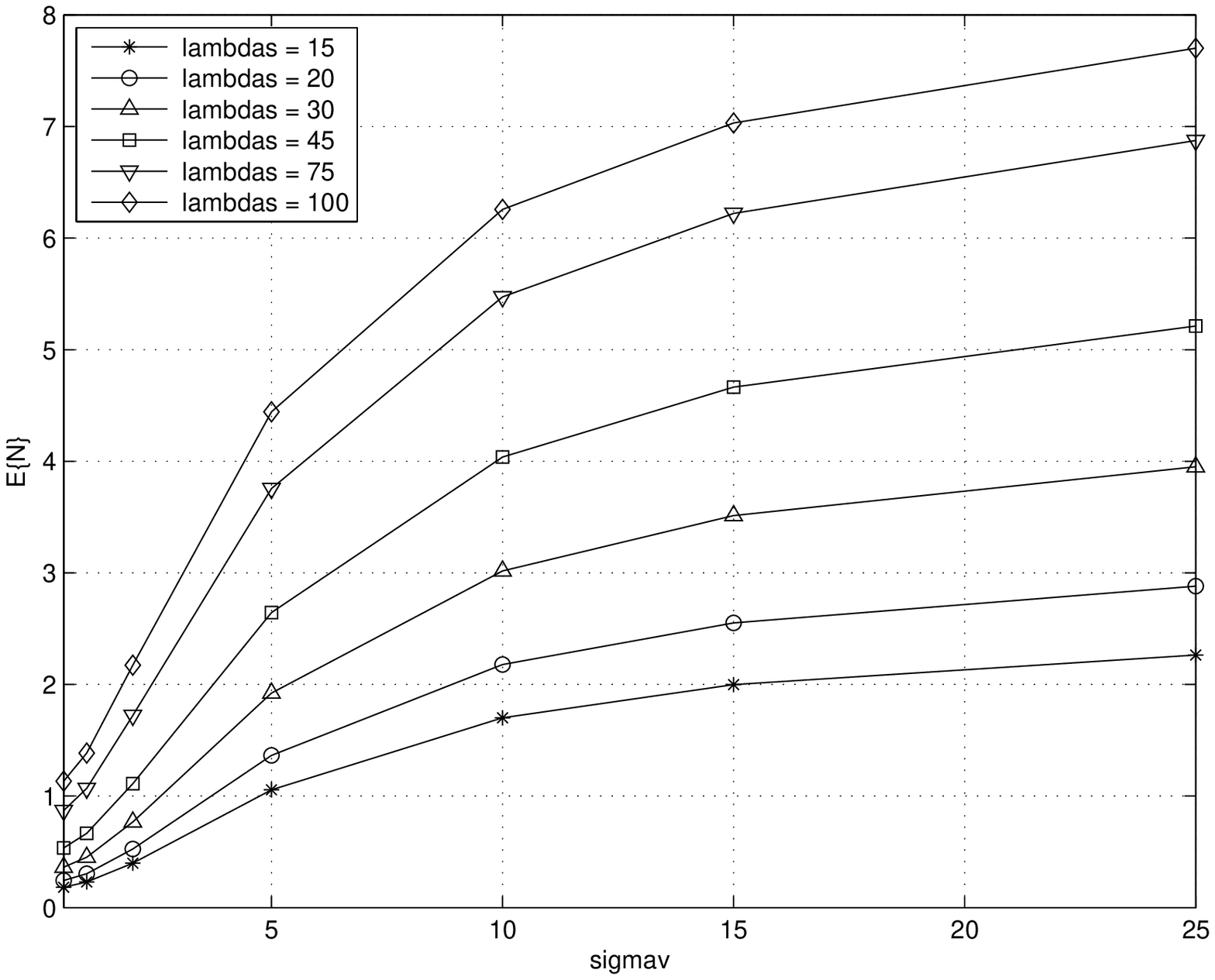}\\
I&II
\end{tabular}
\caption{I. $\EN$ versus the initial density of the sink nodes $\lambda_s$ (number of nodes per 1000 meters), for different values of $\sigma_0$. II. $\EN$ versus $\sigma_0$, for different values of $\lambda_s$ }\label{general_EN_lambdas}
\end{figure}

Figs.~\ref{general_EN_lambdas}-I and \ref{general_EN_lambdas}-II illustrate the numerical values of  $\EN$ in (\ref{simplified}) for different mobility parameters of the sensor node and for different initial density of the sink nodes. The illustrated plots refer to a scenario with a number of sink nodes uniformly distributed in a range of 1000 meters ($D$=500m) on the $x$ axis except for a segment of length 40 meters centered at the origin (where the sensor node is initially located) corresponding to $h=20m$. Each sink node can collect the sensor node information if their distance does not exceed 20 meters ($d$=20m). The sink nodes are all moving with the same mobility parameter $\sigma_j=\sigma;$ for $j=1,2,\ldots\bm_{\rm s}$.

The time instants at which the connectivity of the network is observed have been selected to be 5, 7, 9, ..., 23 seconds and hence the $\bN\leq 10$. As also seen in these figures, the expected number of covered time instants increases with the initial density of the sink nodes $\lambda_s$ and also with the mobility parameter of the sensor node $\sigma_0$.

 The diversity gain introduced by the mobility is evident in these figures. The more \emph{mobile} the network becomes, the larger will be the chance of connectivity in different time instants and collectively, the node will have a larger average number of communication opportunities. This provides a larger number of paths from the sensor to the sink nodes. The signals received over different paths can be combined to detect the transmitted data with a higher reliability. It is also evident from these figures that the curves saturate for large values of $\sigma_0$ to a limit that depends on $\lambda_s$. This limit is actually $2 n d\lambda_s$ as discussed in Section~\ref{largesigma0}. Combining this property, with the exponential relationship between $\Pr\{ \bI_k=1\}$ and $\sigma_0$, and observing the curves of Fig.~\ref{general_EN_lambdas}-II we can approximate (\ref{simplified}) with
 \begin{equation}
   \EN\approx 2n d\lambda_s (1-\displaystyle e^{-\displaystyle\kappa\sigma_0}).
 \end{equation}
 Here, $\kappa$ will be a parameter depending on the sink node mobility parameters. Using this approximation, we can observe that
 \begin{eqnarray}
   \frac{\partial(\EN)}{\partial \sigma_0}\approx 2n d \lambda_s\kappa e^{-\kappa\sigma_0}
 \end{eqnarray}
 which is a decreasing function of $\sigma_0$ justifying the slower growth of $\EN$ for larger $\sigma_0$. For smaller values of $\sigma_0$, $1-e^{-\kappa\sigma_0}$ can be approximated with $\kappa\sigma_0$ leading to $\EN\approx 2nd\kappa\lambda_s\sigma_0$ which is observed as the linear growth with slopes increasing with increasing $\lambda_s$ in Fig.~\ref{general_EN_lambdas}-II.

%

\begin{figure}
\centering
\begin{tabular}{cc}
\psfrag{sigmav}{\tiny $\sigma_0$}
\psfrag{expectation of Nd}{\tiny $\mathbb{E}\{\bN_d\}$}
\psfrag{ lambdas = 0.015}{\tiny $\lambda_s$ = 0.015}
\psfrag{ lambdas = 0.02}{\tiny $\lambda_s$ = 0.02}
\psfrag{ lambdas = 0.03}{\tiny $\lambda_s$ = 0.03}
\psfrag{ lambdas = 0.045}{\tiny $\lambda_s$ = 0.045}
\psfrag{ lambdas = 0.075}{\tiny $\lambda_s$ = 0.075}
\psfrag{ lambdas = 0.1}{\tiny $\lambda_s$ = 0.1}
  \includegraphics[width=85mm]{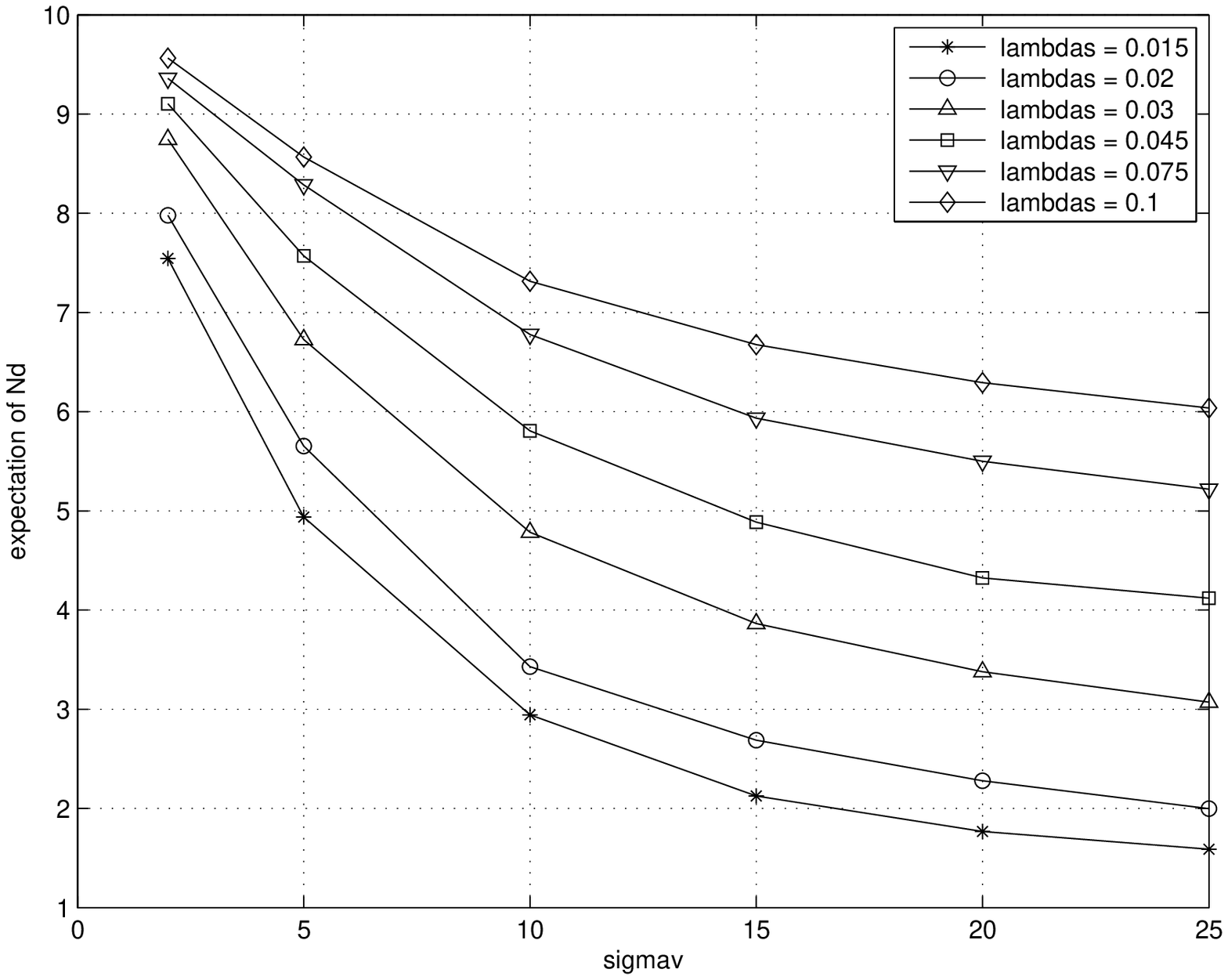}&
  \psfrag{sigmav}{\tiny $\sigma_0$}
  \psfrag{ lambdas = 0.015}{\tiny $\lambda_s$ = 0.015}
\psfrag{ lambdas = 0.02}{\tiny $\lambda_s$ = 0.02}
\psfrag{ lambdas = 0.03}{\tiny $\lambda_s$ = 0.03}
\psfrag{ lambdas = 0.045}{\tiny $\lambda_s$ = 0.045}
\psfrag{ lambdas = 0.075}{\tiny $\lambda_s$ = 0.075}
\psfrag{ lambdas = 0.1}{\tiny $\lambda_s$ = 0.1}
\psfrag{Outage Probability}{\tiny Outage Probability}

  \includegraphics[width=85mm]{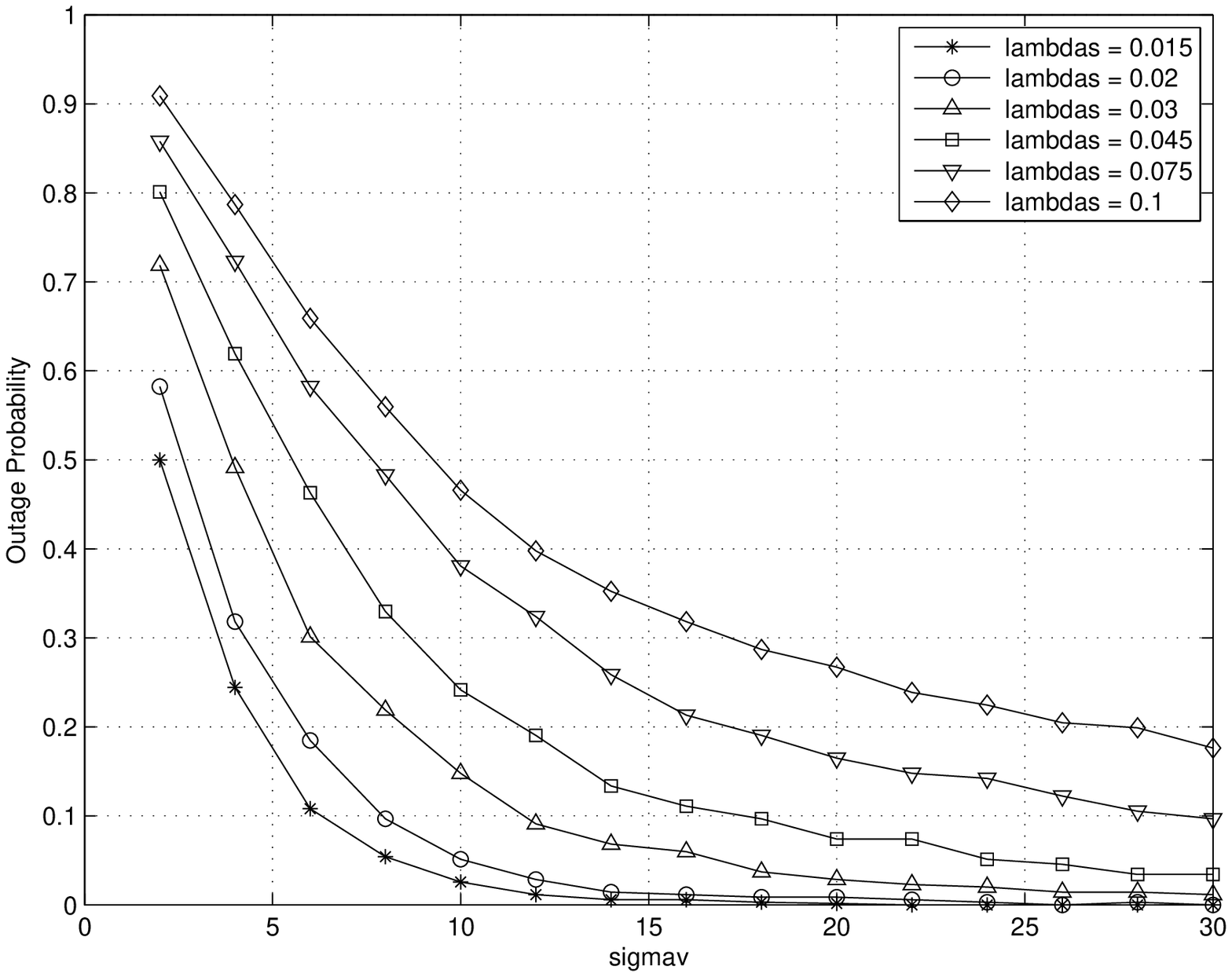}\\
  I&II
\end{tabular}
\caption{I. Expectation of the longest delay $\bN_d$ versus the sensor node mobility parameter, $\sigma_0$, for different values of sink nodes density, $\lambda_s$. II. Probability of outage versus the sensor node mobility parameter, $\sigma_0$, for different values of sink nodes density, $\lambda_s$}\label{N_d}
\end{figure}

Fig.~\ref{N_d}-I illustrates the expected value of the largest number of consecutive non-covered time instants $\bN_d$ versus the mobility parameters of the sensor node. The setup is similar to that used in the numerical evaluation of $\mathbb{E}\{\bN\}$. As evident in Fig.~\ref{N_d}-I, the mobility of the sensor node significantly decreases the expectation of $\bN_d$.

Fig.~\ref{N_d}-II illustrates the probability of outage $\Pr\{\bN=0\}$ versus the sensor mobility parameters $\sigma_0$ for different values of $\lambda_s$. As expected from the mathematical proof, probability of outage decreases with increasing the mobility of the sensor node.

\section{conclusion}\label{conclusion}
In this paper, we introduced the novel concept of \emph{mobility diversity} for mobile sensor or
communication networks as the diversity introduced by transmitting data over different topologies
of the network. We showed how node mobility can provide diversity by changing the topology of the
network and studied a simple network with one-dimensional mobility. More specifically, we
considered a mobile network with a sensor node and a number of sink nodes moving along the $x$-axis
and used a Wiener process mobility model to describe the one-dimensional motion of the nodes. Assuming that the network topology evolves with time and assuming that the connectivity of the sensor node to at least one sink node is needed for successful communication, we calculated three performance measures of the network, i) the expected number of time instants, where the sensor node is connected to at least one sink node, ii) the probability of outage, being the probability that no sink node is in the vicinity of the sensor node in the $n$ observation time instants, and finally iii) the maximum number of consequent failures in the communication.

Our theoretical analysis and numerical experiments show that increasing the mobility parameter
of the sensor node enhances the performance measures of the sensor network in three ways. Increasing mobility, increases the average number of successful transmissions, decreases the probability of outage and decreases the maximum delay that the sensor node will encounter in transmitting its data to the destination.
%
%
%
%

\bibliographystyle{IEEEtran}
\bibliography{IEEEabrv,reference}
\end{document}